\documentclass[journal,twoside,web]{ieeecolor2}
\usepackage{generic}
\usepackage{cite}
\usepackage{amsmath,amssymb,amsfonts}
\usepackage{algorithmic}
\usepackage{graphicx}
\usepackage{textcomp}
\usepackage{booktabs}
\def\BibTeX{{\rm B\kern-.05em{\sc i\kern-.025em b}\kern-.08em
    T\kern-.1667em\lower.7ex\hbox{E}\kern-.125emX}}
\begin{document}
\title{Label-Free Deep-Tissue Peripheral Nerve Detection with a Handheld Multimodal OCT Probe and NerveDetNet}
\author{Yihan Wang, Ruilin You, Shaobai Li, Jiabin Chen, Bofan Song, Anh D. Le, Rongguang Liang
\thanks{This work is supported by NIH (Grant $\#$R01DE030682). Yihan Wang and Ruilin You contributed equally to this work. (Corresponding authors: Rongguang Liang)}
\thanks{Yihan Wang, Ruilin You, Shaobai Li, Jiabin Chen, Bofan Song are with Wyant College of Optical Sciences, University of Arizona, Tucson, Arizona 85721, USA}
\thanks{Anh D. Le is with Department of Oral \& Maxillofacial Surgery, University of Pennsylvania School of Dental Medicine, Philadelphia, PA 19104, USA}
\thanks{Rongguang Liang is with Wyant College of Optical Sciences, University of Arizona, Tucson, Arizona 85721, USA (rliang@optics.arizona.edu)}}

\maketitle

\begin{abstract}
Peripheral nerves buried beneath intact tissue are difficult to visualize during surgery and remain inaccessible to white light wide-field imaging and other surface optical imaging methods. Existing OCT nerve studies have largely relied on exposed nerves or polarization contrast with limited depth penetration, restricting their value for subsurface intraoperative guidance. Here, we introduce, to our knowledge, the first label-free framework for detecting peripheral nerves beneath unopened tissue and resolving their depth using intensity-based OCT structural signatures alone. The framework combines a handheld multimodal probe, integrating swept-source OCT with co-registered white light and autofluorescence imaging, with a ``confirm-then-capture'' workflow designed for practical surgical use. To enable efficient analysis of sparsely sampled OCT volumes, we develop NerveDetNet, a lightweight 2.5D segmentation network that recovers weak and spatially displaced nerve signals by incorporating spatial context, frame-order information, and shift-tolerant correlations across frames through a dedicated nerve feature correlation module. In ex vivo tissue experiments, NerveDetNet consistently outperformed six representative 2D baselines across all frame spacings, achieving a Dice score of 0.725 under the sparsest sampling condition while using approximately half the model parameters. End-to-end validation demonstrated localization of nerves invisible at the surface and depth-resolved detection up to 1.3--1.4~mm below the tissue surface, with OCT derived depth maps overlaid directly onto the surgical view. Together, these results establish a practical label-free approach for subsurface nerve visualization that supports intraoperative compatibility, enables efficient sparse-volume analysis, and provides depth-resolved guidance without tissue opening, contrast agents, or nerve exposure.
\end{abstract}


\section{Introduction}

Iatrogenic injury to peripheral nerves is a significant complication across many surgical procedures, and can lead to chronic pain, sensory loss, or permanent motor deficit~\cite{antoniadis2014iatrogenic,rasulic2017iatrogenic}. Small peripheral nerves are often visually indistinguishable from surrounding muscle, fat, and connective tissue, and may be concealed beneath thin overlying layers. As a result, reliable intraoperative identification is challenging, particularly before a nerve has been surgically exposed. A real-time method to localize peripheral nerves without exogenous labeling or deeper tissue dissection would provide substantial clinical benefit and prevent inadvertent nerve injury.

Existing intraoperative nerve-identification techniques each leave part of this need unmet. Electrophysiological monitoring confirms neural activity but does not image the nerve and offers limited spatial localization. Fluorescent nerve-targeting probes render nerves conspicuous but require exogenous contrast agents and the associated regulatory and toxicity considerations~\cite{whitney2011peptides}. Label-free contrast has therefore been explored across multiple optical modalities. Intrinsic nerve autofluorescence has been investigated for intraoperative nerve identification in ex vivo, animal, and clinical settings~\cite{dip2021nerveAF,dip2022nuvInVivo}. In addition, diffuse-reflectance and multimodal optical systems have been used to visualize nerves and adjacent vasculature in real time during dissection~\cite{cha2018multimodal,throckmorton2023drs,haugen2023opticalprops}. These methods, however, are fundamentally surface techniques: like white light, they visualize only nerves already exposed at the surgical surface. Detecting nerves embedded beneath overlying tissue remains a persistent and widely acknowledged challenge in this field~\cite{haugen2023opticalprops}.

Optical coherence tomography (OCT) is, in principle, well suited to this gap, because it provides label-free, micrometer-scale, depth-resolved imaging and penetrates a short distance into scattering tissue. Its most established use for neural imaging, however, has been in the eye, where the optic nerve head and retinal nerve fiber layer are imaged through transparent ocular media~\cite{shin2024opticnerve}. In opaque peripheral tissue, the situation differs, and reported OCT studies of peripheral nerves have largely imaged nerves that were already surgically exposed or excised~\cite{cha2018multimodal,psoct_vagus}. A prominent line of work incorporates polarization sensitivity (PS-OCT), exploiting the birefringence of myelinated fascicles to enhance nerve-to-background contrast and to assess fascicular anatomy, injury, or stimulation safety~\cite{psoct_vagus,psoct_crinjury,psoct_electrostim,liao2025polarimetric}. While powerful, polarization contrast is intrinsically depth-limited: PS-OCT of full-thickness nerve has been reported as constrained by tissue-limited imaging depth~\cite{psoct_vagus}, and in deeper tissue, increasing multiple scattering degrades the degree of polarization and corrupts birefringence and retardance estimates~\cite{psoct_dopu_depth}. As a result, polarization-based discrimination weakens precisely where a nerve is most deeply buried and most vulnerable to inadvertent injury. Across these approaches, the common limitation is that nerves are imaged after exposure, rather than detected while still covered, which is the situation in which injury actually occurs.

In this work, we demonstrate a compact-handheld multimodal OCT system for subsurface peripheral nerve detection, together with NerveDetNet, a sparse frame reconstruction and segmentation framework designed for efficient depth resolved nerve localization. Rather than enhancing nerves that are already exposed or visible, our approach targets the more challenging problem of identifying nerves concealed beneath intact tissue directly from their intrinsic intensity based OCT structural signatures. This is enabled by the characteristic Bands of Fontana, a quasi periodic banded texture arising from the undulating organization of nerve fiber bundles, which can be captured by conventional OCT without labels, contrast agents, or polarization-based enhancements. To make this capability practical for intraoperative use, the system integrates OCT with co registered white light and autofluorescence imaging in a freehand probe, enabling detected nerve locations and depths to be overlaid onto the surgical view. At the same time, NerveDetNet addresses the computational challenge of sparse volume analysis by recovering weak, depth attenuated, and spatially displaced nerve signals across frames using spatial context, frame order cues, and shift tolerant feature correlations. Together, the integrated system and algorithm provide a practical route for label-free, depth resolved visualization of buried peripheral nerves during surgery.

The contributions of this work can be summarized as follows:
\begin{itemize}
\item A handheld multimodal OCT probe that integrates swept-source OCT with co-registered white light and autofluorescence imaging and a confirm-then-capture real-time workflow, enabling freehand, motion-free acquisition and pixel-level overlay of detected nerves onto the surgical field.
\item NerveDetNet, a sparse frame 2.5D segmentation network whose nerve feature correlation module (NFCM) guided decoder integrates spatial, order, and shift-tolerant cross-frame correlation cues to recover weak and displaced nerve signals under sparse sampling, outperforming in detection while keeping inference cost close to other 2D neural network segmentation frameworks.
\item To our knowledge, the first demonstration of detecting and depth-resolving peripheral nerves beneath intact, unopened tissue directly from the intensity-based OCT signature—rather than imaging already-exposed nerves—at depths reaching 1.3--1.4~mm below the surface.
\end{itemize}
We describe the probe and detection workflow of the nerve in Sec.~\ref{sec:method}, the experimental setup in Sec.~\ref{sec:experiments}, and the segmentation and end-to-end results in Sec.~\ref{sec:results}; we then discuss the limitations and potential impact in Sec.~\ref{sec:discussion}.

\section{method}
\label{sec:method}

\subsection{Handheld multimodal OCT probe}

\begin{figure}[htbp]
    \centering
    \includegraphics[width=\columnwidth]{ 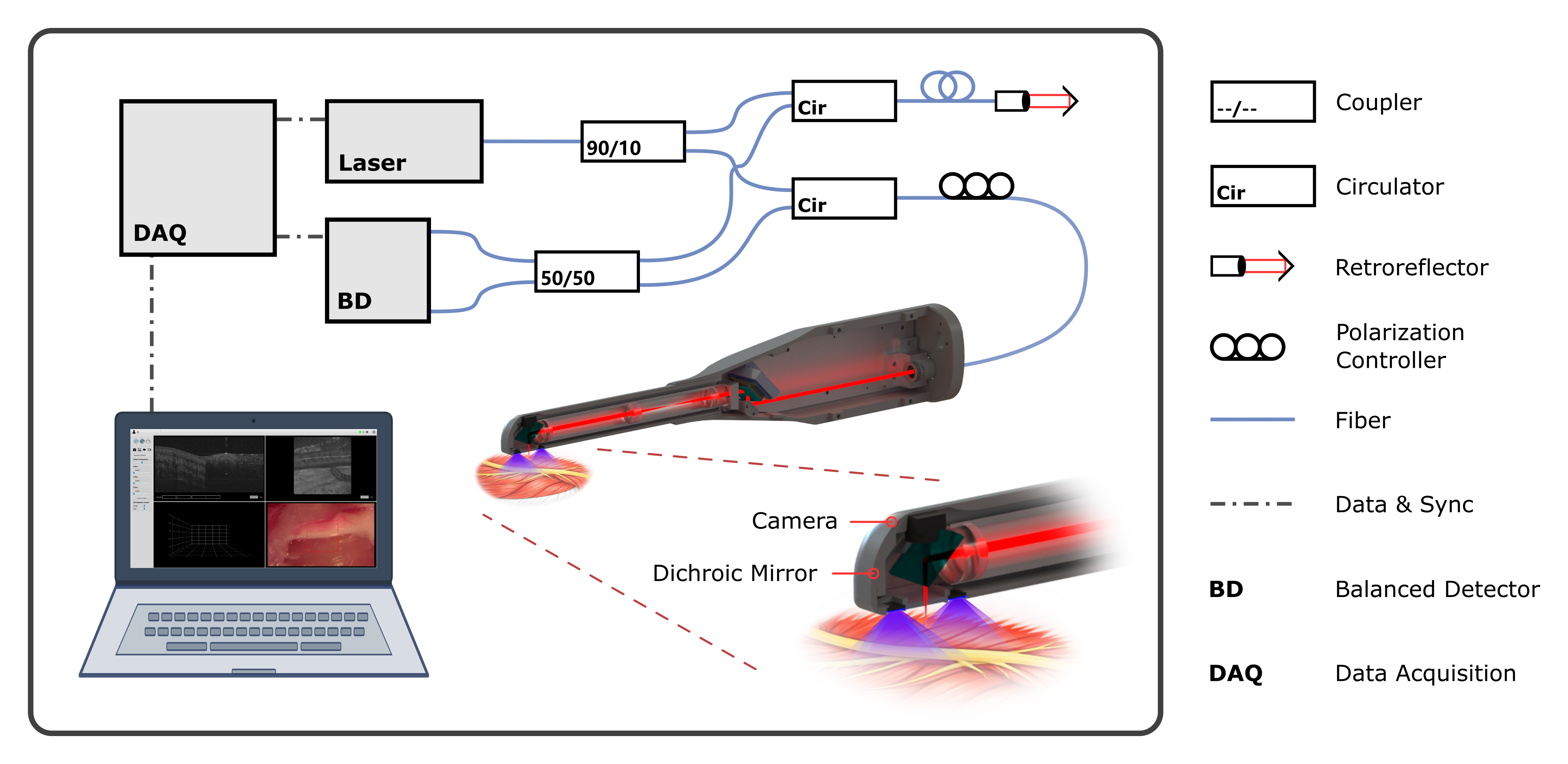}
    \caption{Schematic of the optical system of the handheld multimodal OCT probe. 
    }
    \label{fig:octsystem}
\end{figure}

The handheld multimodal OCT probe (Fig.~\ref{fig:octsystem}) detects peripheral nerves beneath intact tissue by integrating structural OCT with co-registered white light and autofluorescence imaging. The swept-source OCT operates at a 1300~nm center wavelength using a MEMS-VCSEL source (SVM132-3023, Thorlabs), providing a 100~nm tuning range (at $-15$~dB), a 200~kHz sweep rate, $>$100~mm coherence length, and 30~mW output. A built-in Mach--Zehnder interferometer (32~mm optical path difference) generates the k-clock with 2048 samples per sweep, and a balanced photodetector pair (0.85~A/W, 0.1--500~MHz) suppresses common-mode noise. A custom set of three cemented doublet lenses yields an image-space NA of 0.04 and a $6\times6$~mm field of view. A 10~mm working distance avoids sample contact, and a telecentric configuration eliminates perspective-induced lateral magnification variation. The theoretical axial and lateral resolutions reach 7.5~$\mu$m and 21~$\mu$m in air, and the measured penetration depth in muscle is up to 1.8~mm.

For multimodal imaging, a dichroic mirror coaxially aligns the optical axes of the camera and the OCT beam, with a linear polarizer and a UV cut-off filter in front of the camera to suppress specular reflection and autofluorescence stray light. The probe head incorporates four white light LEDs for illumination, each fronted by a polarizer crossed with the camera's, and four ultraviolet LEDs with a 425~nm short-pass filter for autofluorescence excitation. The camera provides a 1080p stream at 30~fps over a $30.5\times17$~mm field of view for guidance. The distal extension formed by the camera and OCT lens group ($18\times18\times100$~mm) allows imaging in confined regions.

The proximal handle houses a MEMS scanning mirror (Mirrorcle Technologies), actuated through a multifunction I/O board (PCIe-6738, National Instruments) that also controls the LED driver and handle shortcut buttons. The OCT signal is digitized by a high-speed acquisition card (ATS9371, Alazar Technologies; 12-bit, 1.2~GS/s) and streamed to a laptop for real-time CUDA processing and display.

\subsection{Live-view nerve detection workflow}

\begin{figure*}[htbp]
    \centering
    \includegraphics[width=12.5cm]{ 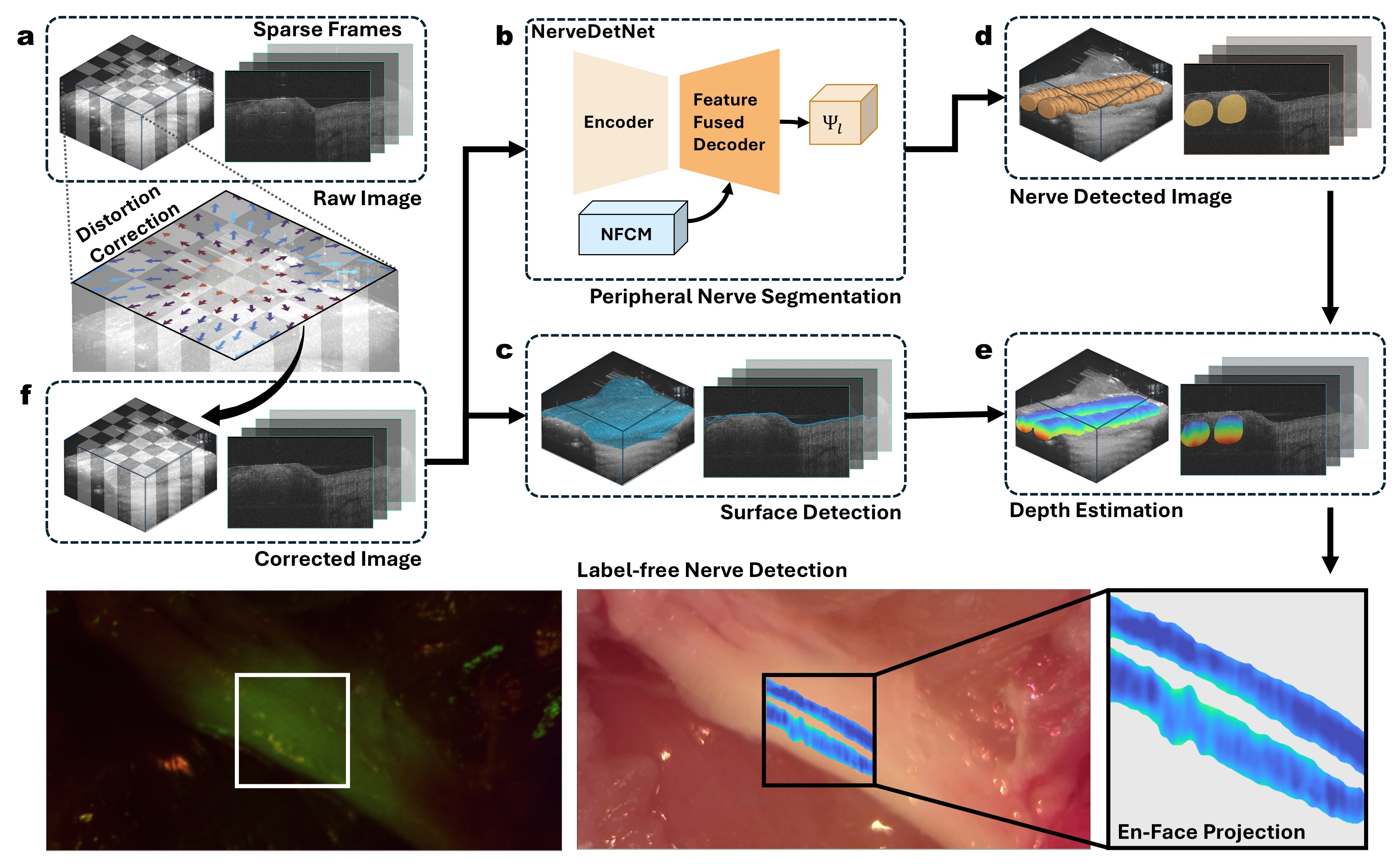}
    \caption{Pipeline of multimodal OCT peripheral nerve detection.
    \textbf{(a)} Distortions are corrected via a calibrated displacement field.
    \textbf{(b)} NerveDetNet segments peripheral nerves from the corrected volumes or sparse frames.
    \textbf{(c)} In parallel, the tissue surface is detected.
    \textbf{(d)} The nerve detected volume with the segmented region overlaid.
    \textbf{(e)} Combining the nerve region with the tissue surface yields the subsurface nerve depth map.
    \textbf{(f)} Results for visualization and cross-validation: autofluorescence (left), white light fused with the OCT-derived depth map (center and right).}
    \label{fig:workflow}
\end{figure*}

The system supports a full volumetric scan ($512\times512$ A-lines, 2~s) and a sparse-frame mode (512 A-lines per frame, $>$250~fps). After acquisition, digital dispersion correction is performed in the spectral domain using the single arbitrary measurement of the mirror-reflection (SAMMR) method~\cite{wang2023sammr}, which extracts the dispersion-induced phase delay directly without fitting dispersion coefficients. From a single mirror-reflection measurement, the complex interference signal is obtained by Hilbert transform, and its phase $\phi(k) = 2k\,z_{\mathrm{OPD}} + \phi_D(k)$ comprises an optical-path-difference term and the dispersion phase $\phi_D(k)$. After locating $z_{\mathrm{OPD}}$ from the PSF peak, $\phi_D(k)$ is isolated and, prior to the Fourier transform, compensated into each interference signal via a CUDA kernel for real-time dispersion compensation.

As shown in Fig.~\ref{fig:workflow}(a), lateral distortion correction is then applied. Because MEMS scanning produces distortion that is neither radially symmetric nor well described by parametric barrel/pincushion models, we adopt a non-rigid correction based on an equally spaced dot-array calibration target. In the calibration en-face image, dot centroids are detected with a Difference-of-Gaussians operator ($\sigma_s=1.5$, $\sigma_l=6.0$) and Otsu thresholding; a stray dot occasionally present at the center of a four-dot cell has four nearest neighbors at $s/\sqrt{2}$ (where $s$ is the grid spacing) and is rejected at $\approx 0.854\,s$. The remaining dots are indexed by a four-neighbor breadth-first traversal, establishing a one-to-one correspondence between the detected coordinates $\mathbf{p}_i$ and the ideal, equally spaced coordinates $\mathbf{q}_i$.

A thin-plate spline (TPS) is then fitted to map ideal to distorted coordinates:
\begin{equation}
f(\mathbf{x}) = a_0 + \mathbf{a}^\top \mathbf{x} + \sum_{i=1}^{N} w_i\, \varphi\big(\|\mathbf{x} - \mathbf{q}_i\|\big), \quad \varphi(r) = r^2 \log r,
\end{equation}
with coefficients obtained by minimizing the fitting residual together with a regularized bending-energy term weighted by $\lambda=0.3$. Because the TPS assumes no symmetry of the distortion, it captures asymmetric deformations in which the two lateral axes are distorted to different degrees. The corrected image is resampled onto a fixed $512\times512$ grid by bicubic interpolation. Since the telecentric configuration keeps lateral magnification essentially invariant with depth, the same mapping is applied to every depth slice, giving the corrected en-face frame and every B-scan a consistent lateral scale for multimodal registration.

The corrected image is then processed along two paths. On one, it is fed into the pre-trained NerveDetNet (Fig.~\ref{fig:workflow}(b)) for peripheral nerve segmentation, yielding the nerve distribution region (Fig.~\ref{fig:workflow}(d)). On the other, tissue-surface detection is performed along individual A-lines (Fig.~\ref{fig:workflow}(c)). Each B-scan is median-filtered along depth and the axial gradient $G = \partial \tilde{I}/\partial z$ is computed by central differencing, with the top $m=10$ rows zeroed to remove the zero-optical-path artifact; the surface $\hat{z}_{\mathrm{surf}}(x)$ is taken as the location of maximum positive gradient. Columns with a weak peak ($p(x) < \kappa\,\eta$, $\eta=\mathrm{median}\,|G|$, $\kappa=25$) fall back to a local Otsu threshold. The per-frame profiles are then outlier-rejected, interpolated, smoothed, and median-filtered across adjacent B-scans to yield a coherent tissue-surface map $z_{\mathrm{surf}}(x,y)$.

Combining the tissue surface with the nerve region, and converting the optical path difference with a group refractive index $n \approx 1.4$, gives the nerve depth and distribution beneath the tissue (Fig.~\ref{fig:workflow}(e)). The white light camera and the OCT field of view are co-registered at the pixel level by a one-time calibration combining hardware optical-axis alignment via the dichroic mirror with a software digital offset correction, so the OCT derived depth map overlays directly onto the white light image (center of Fig.~\ref{fig:workflow}(f)) without per-acquisition registration. The autofluorescence image (left of Fig.~\ref{fig:workflow}(f)) is displayed concurrently as an independent cross-reference for the operator.

\subsection{OCT Nerve Feature Characteristics}

\label{sec:nervefeature}

\begin{figure}[hbtp]
    \centering
    \includegraphics[width=8cm]{ 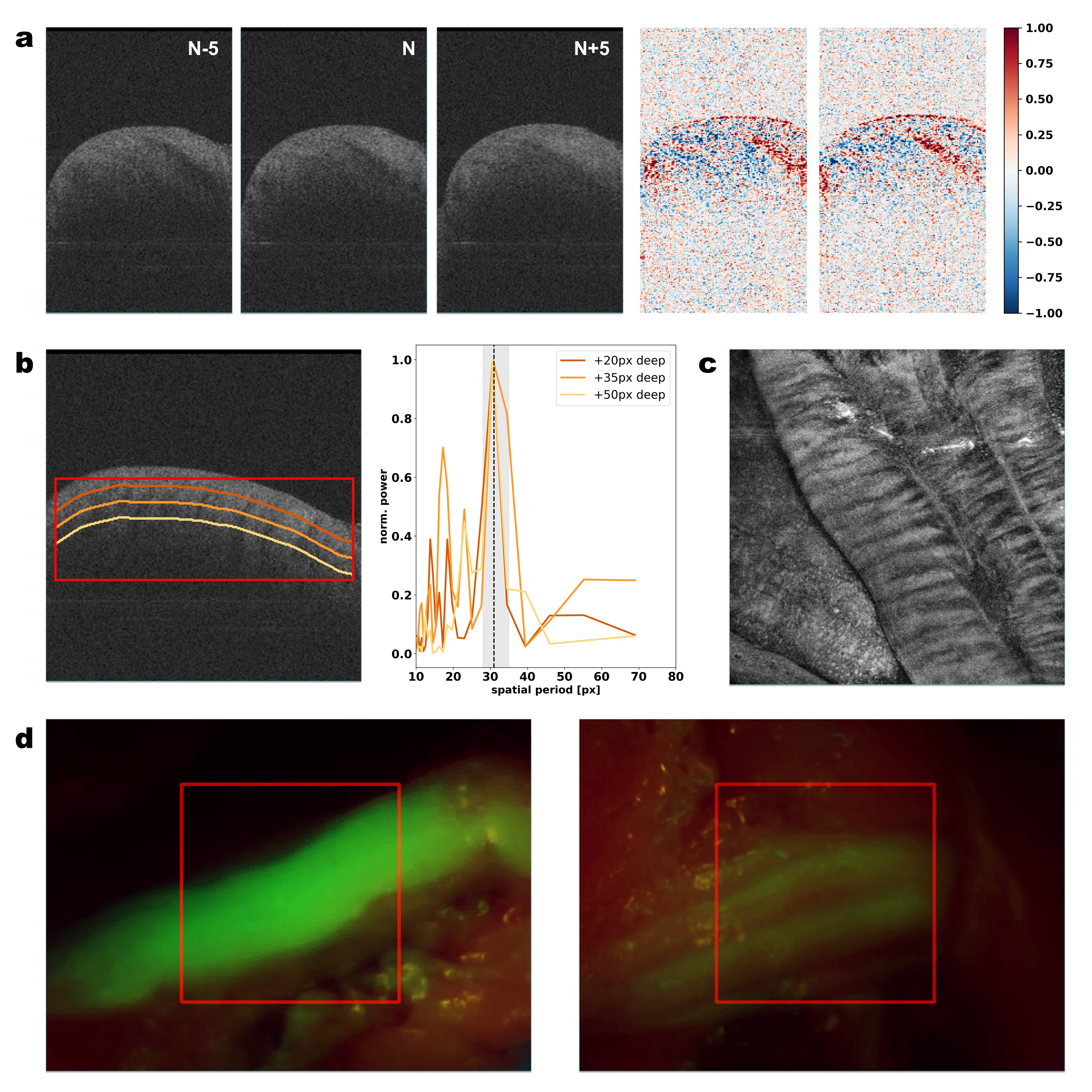}
\caption{OCT signatures and spatial coherence of peripheral nerve.
(a) Surface-registered B-scans separated by 5 frames and corresponding normalized difference maps.
(b) Long-axis intensity profiles showing a shared spatial-frequency peak within the Bands-of-Fontana period range.
(c) CLAHE-enhanced en-face projection showing periodic nerve texture.
(d) Autofluorescence of exposed nerve under \(365\)~nm excitation.}
    \label{fig:nervefeature}
\end{figure}

In OCT, peripheral nerve exhibits the ``Bands of Fontana'', an internal, approximately periodic pattern of bright and dark bands produced by the undulating arrangement of nerve fiber bundles. This texture helps distinguish nerve from the fibrous appearance of muscle and the more diffuse appearance of connective tissue and fat.

The banded structure is most evident in the en-face plane. In the CLAHE-enhanced mean-intensity projection in Fig.~\ref{fig:nervefeature}(c), three nerves are identifiable by periodic banded texture that is absent in surrounding tissue. To quantify this periodicity, long-axis intensity profiles were sampled from the central nerve at three depths (Fig.~\ref{fig:nervefeature}(b)). Their power spectra showed a shared peak at a spatial period of \(\sim31\)~px across depth, whereas muscle profiles showed no comparable peak. This frequency consistency across depth motivates the NFCM spatial guidance group.

Because nerves are continuous three dimensional structures, their banded pattern evolves coherently across adjacent B scans, unlike speckle. Fig.~\ref{fig:nervefeature}(a) shows three B scans through the central nerve, separated by 5 frames. After registration by the tissue surface contour, normalized differences between successive frames show a banded pattern shifting as a coherent unit rather than changing randomly, indicating ordered low frequency evolution along the volume direction with limited displacement between frames. These properties motivate the NFCM order and shift correlation guidance groups. They also explain why detection becomes harder as frame spacing \(SF\) increases: nerves can shift substantially between sparsely sampled frames, making explicit shift tolerant correlation increasingly important, which is the regime where NerveDetNet shows its largest advantage. Video~1 further illustrates nerve morphology through sequential B scans, volumetric rendering, and multi axis cross sections.

Peripheral nerve also shows intrinsic autofluorescence, which has been used for nerve visualization without exogenous labeling~\cite{dip2021nerveAF}. Consistent with prior reports, fresh exposed nerve fluoresced more strongly than muscle under 365~nm excitation (Fig.~\ref{fig:nervefeature}(d)). However, this contrast is sensitive to sample freshness and can be confounded by adipose tissue, which also fluoresces; in our commercially sourced samples it was sometimes less pronounced. We therefore used autofluorescence only as an auxiliary cross reference and based detection on the OCT structural signature, which remains available through overlying tissue and does not depend on fluorescence strength.

\subsection{NerveDetNet Architecture}
\label{sec:nervedetnet}

\begin{figure}[htbp]
    \centering
    \includegraphics[width=\columnwidth]{ 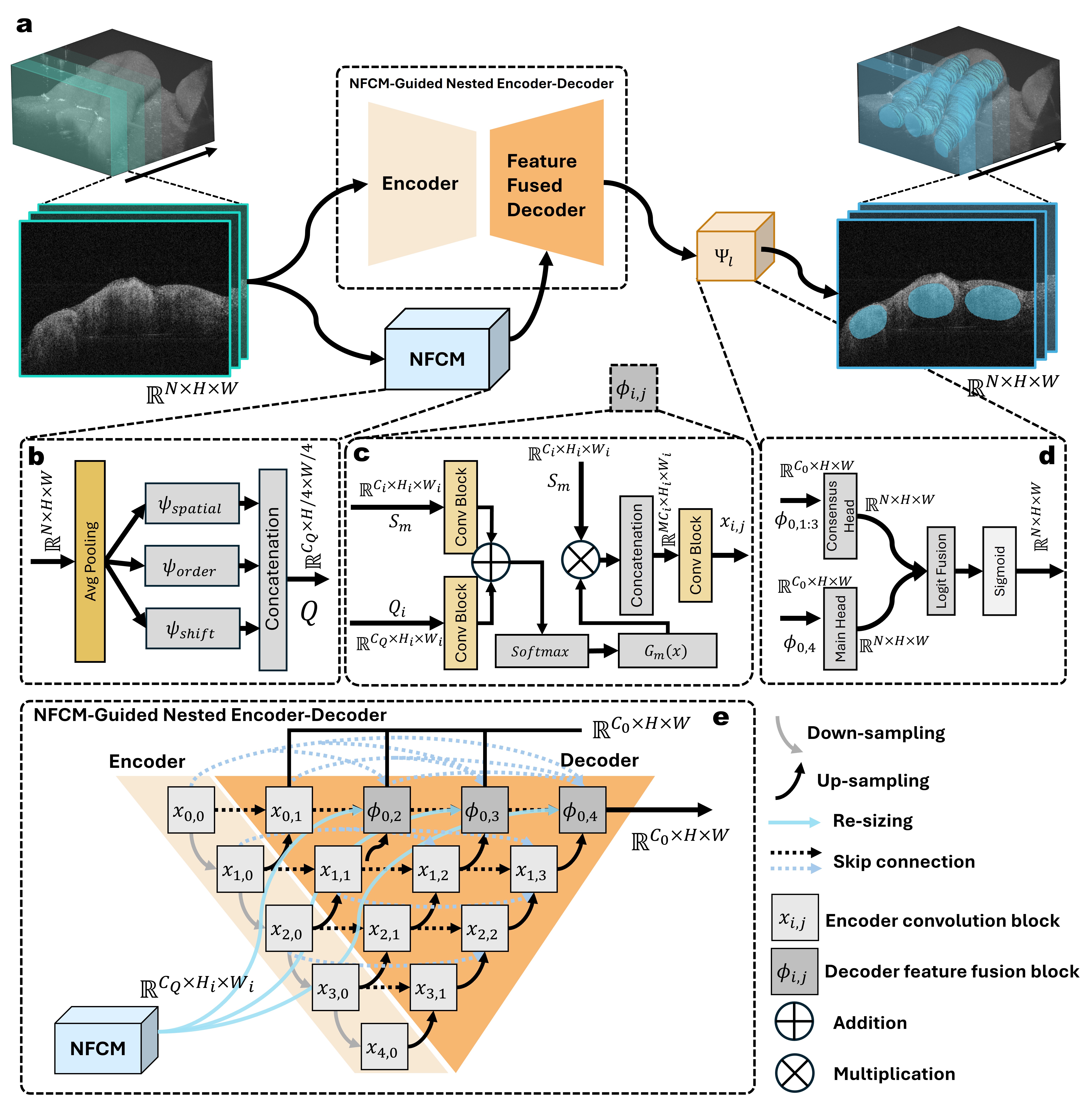}
    \caption{Overview of NerveDetNet for sparse-frame OCT nerve detection.
    (a) Sparse B-scan stack and frame-wise prediction.
    (b) NFCM guidance extraction.
    (c) NFCM-guided decoder fusion.
    (d) Main and consensus logit fusion.
    (e) Compact nested encoder--decoder topology with guidance applied along the final high-resolution path.}
    \label{fig:NerveDetNet}
\end{figure}

NerveDetNet formulates subsurface OCT nerve detection as sparse frame 2.5D segmentation. For each sample, the input is an ordered stack \(\mathbf{X}=[X_1,\ldots,X_N]\in\mathbb{R}^{N\times H\times W}\), where \(N=5\), \(H\) and \(W\) denote the axial and lateral dimensions, and adjacent frames are separated by spacing \(SF\). The network predicts \(\hat{\mathbf{Y}}=f_{\theta}(\mathbf{X})\in[0,1]^{N\times H\times W}\), where each output slice is the nerve probability map of the corresponding input B-scan. This formulation preserves strong in-plane OCT morphology and limited through-volume context while avoiding the memory and latency cost of full 3D convolutional segmentation~\cite{cicek2016threedunet,milletari2016vnet,zhang2022bridging2d3d}.

As shown in Fig.~\ref{fig:NerveDetNet}, NerveDetNet uses a U-Net++-style nested encoder--decoder~\cite{zhou2020unetpp} with OCT-specific NFCM guidance. Encoder features are computed as \(x_{0,0}=E_0(\mathbf{X})\) and \(x_{i,0}=E_i(x_{i-1,0})\), \(i=1,\ldots,4\). A standard nested decoder node is
\begin{equation}
\begin{aligned}
x_{i,j}=D_{i,j}\!\Big(
\operatorname{Concat}\!\big[x_{i,0},\ldots,x_{i,j-1},U(x_{i+1,j-1})
\big]\Big),
\end{aligned}
\label{eq:standard_nested_decoder}
\end{equation}
where \(D_{i,j}\) is a convolutional decoder block, \(U\) denotes upsampling, and \(\operatorname{Concat}[\cdot]\) denotes channel-wise concatenation.

\subsubsection{Nerve Feature Correlation Module}

The NFCM computes compact guidance maps from the input stack at quarter resolution, \(\mathbf{X}^{(q)}=A_4(\mathbf{X})\in\mathbb{R}^{N\times H/4\times W/4}\), where \(A_4\) denotes spatial average pooling with stride 4. The final NFCM concatenates three cue groups:
\begin{equation}
\begin{aligned}
\mathbf{Q}
=&\,\operatorname{Concat}\!\big[
\psi_{\mathrm{spatial}}(\mathbf{X}^{(q)}),
\psi_{\mathrm{order}}(\mathbf{X}^{(q)}),\psi_{\mathrm{shift}}(\mathbf{X}^{(q)})
\big],\\
\mathbf{Q}
\in&\,\mathbb{R}^{C_Q\times H/4\times W/4}.
\end{aligned}
\label{eq:nfcm_guidance}
\end{equation}
These groups encode local OCT structure, frame order, and shift-tolerant similarity between neighboring sampled frames. The shift-correlation cue helps preserve coherent nerve morphology despite apparent inter frame displacement while reducing sensitivity to isolated speckle and unrelated high-frequency tissue patterns. Summary and coordinate guidance groups were excluded from the final implementation because they did not improve validation performance.

\subsubsection{NFCM-Guided Decoder Fusion}

For decoder node \((i,j)\), let \(\mathcal{S}_{i,j}=\{S_m\}_{m=1}^{M}\) denote the incoming source features, including same-resolution features from previous decoder nodes and the upsampled feature from the next deeper level. The NFCM map is resized to the node resolution as \(Q_i=R_i(\mathbf{Q})\in\mathbb{R}^{C_Q\times H_i\times W_i}\). Source scores, source-wise attention, and identity-initialized gates are computed as
\begin{equation}
\begin{aligned}
\ell_m &= g_s(S_m)+[g_q(Q_i)]_m,\alpha_m =
\frac{\exp(\ell_m)}
{\sum_{k=1}^{M}\exp(\ell_k)},\\
G_m &= 1+\gamma(M\alpha_m-1),
\end{aligned}
\label{eq:nfcm_gate_compact}
\end{equation}
where \(g_s\) maps each source feature to a one-channel score map, \(g_q(Q_i)\in\mathbb{R}^{M\times H_i\times W_i}\) produces source-specific NFCM guidance scores, and \(\gamma\) is a learnable gate-strength coefficient initialized to zero. Because the NFCM guidance term is source-specific, it does not act as a common bias and therefore remains effective after source-wise softmax normalization. The initialization gives \(G_m=1\), and uniform attention also preserves an identity gate when \(\gamma=1\). The gated feature is \(\widetilde{S}_m=G_m\odot S_m\), where \(G_m\) is broadcast across the channels of \(S_m\), and the NFCM-guided decoder node is

\begin{equation}
\begin{aligned}
\phi_{i,j}
=D_{i,j}\!\Big(
\operatorname{Concat}\!\big[
&\widetilde{S}_1,\ldots\widetilde{S}_M
\big]\Big).
\end{aligned}
\label{eq:nfcm_fusion}
\end{equation}
Here, \(\odot\) denotes element-wise multiplication. Since \(G_m\) is a single-channel spatial gate, it is applied identically to all channels of \(S_m\). This allows each decoder node to locally balance shallow boundary detail, deeper semantic context, and NFCM-derived cross-frame evidence.

For efficiency, NFCM-guided fusion is applied only along the final high-resolution decoder path, \(\Phi_{\mathrm{path}}=\{\phi_{0,2},\phi_{0,3},\phi_{0,4}\}\), where \(\phi_{0,4}\) is the terminal decoder feature used for main-logit prediction. All other decoder nodes remain standard convolutional decoder blocks.

\subsubsection{Output Heads and Logit Fusion}

The terminal guided feature \(\phi_{0,4}\in\mathbb{R}^{C_0\times H\times W}\) produces the main logits \(L_{\mathrm{main}}=h_{\mathrm{main}}(\phi_{0,4})\in\mathbb{R}^{N\times H\times W}\). A consensus head combines intermediate high-resolution logits as
\begin{equation}
\begin{aligned}
L_{\mathrm{cons}}
&=\sum_{r=1}^{3}\beta_r h_r(z_r),\boldsymbol{\beta}=\operatorname{softmax}(\boldsymbol{\eta}),\\
(z_1,z_2,z_3)
&=(x_{0,1},\phi_{0,2},\phi_{0,3}).
\end{aligned}
\label{eq:consensus_head}
\end{equation}
The final prediction is
\begin{equation}
\begin{aligned}
\hat{\mathbf{Y}}
&=\sigma\!\left(\Psi_l(L_{\mathrm{main}},L_{\mathrm{cons}})\right),\hat{\mathbf{Y}}\in[0,1]^{N\times H\times W}.
\end{aligned}
\label{eq:final_prediction}
\end{equation}

Thus, NerveDetNet remains a stacked-frame 2D segmentation network but explicitly introduces ordered and shift-tolerant cross-frame OCT cues. Because the NFCM operates at \(H/4\times W/4\) resolution and is injected only along the final high-resolution decoder path, it adds cross-frame guidance with limited computational and memory cost and does not require a pretrained teacher, registration network, or separate post-processing model.

\subsection{sparse frame Volume Reconstruction Post-processing}
\label{sec:sparse_volume_reconstruction}

After training, dense OCT probability volumes are reconstructed from sparse frame predictions using a deterministic post-processing procedure. Because predicting every B-scan would require repeated evaluation of local \(N\)-frame stacks, the trained network is first applied only to anchor slices. Let \(\mathbf{V}\in\mathbb{R}^{D\times H\times W}\) denote an OCT volume with \(D\) B-scans. For frame spacing \(SF\), the anchor stride is \(s=SF+1\), and the anchor set is \(\mathcal{A}=\{0,s,2s,\ldots\}\cup\{D-1\}\). For each anchor \(z_a\in\mathcal{A}\), an ordered stack is sampled as
\begin{equation}
\begin{aligned}
I_r(z_a)
&=\operatorname{clip}\!\left(z_a+(r-c)s,0,D-1\right),\\
\mathbf{X}_{z_a}
&=[V_{I_0(z_a)},\ldots,V_{I_{N-1}(z_a)}],
\end{aligned}
\label{eq:anchor_stack}
\end{equation}
where \(r=0,\ldots,N-1\), \(c=\lfloor N/2\rfloor\), and \(\mathbf{X}_{z_a}\in\mathbb{R}^{N\times H\times W}\). The center output channel is assigned to the anchor slice, \(P_{\mathcal{A}}(z_a)=f_{\theta}(\mathbf{X}_{z_a})_c\), yielding sparse anchor probability maps without evaluating the model at every B-scan.

The sparse predictions are converted to a dense probability volume by interpolation along the B-scan direction. Following standard medical-image interpolation practice~\cite{lehmann1999survey,thevenaz2000interpolation}, interpolation is performed on continuous probability maps before thresholding. For a non-anchor slice \(z\) between neighboring anchors \(z_l\) and \(z_r\),
\begin{equation}
\begin{aligned}
P(z)
&=(1-\lambda)P_{\mathcal{A}}(z_l)
+\lambda P_{\mathcal{A}}(z_r),\space \lambda=\frac{z-z_l}{z_r-z_l}.
\end{aligned}
\label{eq:volume_interpolation}
\end{equation}
Anchor slices retain their direct predictions, \(P(z_a)=P_{\mathcal{A}}(z_a)\), and boundary slices outside the first or last anchor interval use the nearest available anchor prediction. This produces a dense probability volume \(\mathbf{P}\in[0,1]^{D\times H\times W}\).

The final binary nerve mask is obtained by thresholding:
\begin{equation}
\hat{\mathbf{M}}(z,h,w)
=
\mathbb{I}\!\left[\mathbf{P}(z,h,w)\geq\tau\right],
\label{eq:mask_threshold}
\end{equation}
where \(\tau=0.5\) unless otherwise specified. This post-processing step contains no trainable parameters, is applied identically to all sparse frame models in the benchmark, and reconstructs a full-resolution volumetric mask while preserving the original OCT volume dimensions.

\section{Experiments}
\label{sec:experiments}
\subsection{Data Acquisition Using a Multimodal Handheld OCT Probe}

Peripheral nerve imaging data were acquired from fresh \textit{ex vivo} chicken thigh specimens purchased from a local grocery store. This food-grade model contains the sciatic nerve embedded within muscle, connective tissue, and fat, providing a low-cost, anatomically heterogeneous platform for label-free nerve detection and a widely used microsurgical training model that avoids live-animal procedures~\cite{jeong2013microsurgical,chen2014novel}. Each thigh was placed inner-surface up at room temperature; a small incision exposed the underlying tissue, and muscle was separated from the femoral bone to identify the sciatic nerve and adjacent vessels. Under bright illumination, the exposed nerve was identified by its Bands of Fontana, providing an independent anatomical reference for subsequent OCT scanning and manual annotation. Because white-light imaging is surface-limited whereas OCT can resolve structural nerve signatures beneath superficial tissue, the dataset was used to evaluate OCT-based subsurface nerve detection.

Imaging was performed with the handheld probe (Sec.~II-A), held freehand at a \(10\pm1\)~mm OCT working distance without contacting the sample, using a confirm-then-capture workflow. After each \(512\times512\) OCT volume scan, an axial-mean en-face projection was reconstructed on the fly using the CUDA pipeline, allowing the operator to screen for motion from en-face continuity. If the volume was motion-free, a handle button triggered co-posed white-light and autofluorescence capture, separated by \(<0.2\)~s; the combined exposure sequence took \(0.5\)~s. Motivated by reported intrinsic nerve autofluorescence~\cite{dip2021nerveAF}, autofluorescence was acquired only as a complementary cross-reference and was not used as a primary detection input because its contrast and adipose confounding are discussed in Sec.~\ref{sec:nervefeature}.

All data were acquired as \(512\times512\) volume scans, and sparse-frame samples were generated by retrospective subsampling (Sec.~\ref{sec:dataset_preparation}). Acquisition and processing ran on a laptop with an Intel Core i7-13800H CPU, NVIDIA RTX~4080 Laptop GPU, and 64~GB RAM, processing each B-scan in real time without backlog. Data were collected from 16 chicken thighs, with each volume acquired at a different site, yielding 22 nerve-positive and 3 nerve-absent volumes. Consistent with Sec.~II-A, OCT retained usable signal down to \(1.8\)~mm in muscle, defining the effective imaging depth for subsurface detection.

\subsection{OCT Volume Dataset and Sparse-Frame Preparation}
\label{sec:dataset_preparation}

The acquired OCT volumes were curated into a sparse-frame segmentation dataset consisting of 25 volumes: 22 nerve-positive volumes with manually annotated binary nerve masks and 3 nerve-absent volumes used as false-positive controls. To reduce acquisition-related intensity bias, source-specific normalization was applied. For each source directory, the mean and standard deviation were computed from the OCT volumes, and each volume was converted to a z-score representation, clipped to \(\pm3\) standard deviations, and mapped to an 8-bit intensity range. The normalization statistics were stored with the dataset metadata and reused during inference when source-specific normalization was enabled.

The nerve-absent volumes were included to provide explicit examples in which no nerve was present in the scanned field, rather than to compensate for foreground--background class imbalance, since the nerve-positive volumes already contained substantial non-nerve tissue. These negative-control volumes helped the models learn volume-level non-nerve appearances and reduce false-positive predictions in nerve-absent scans.

Sparse-frame samples were generated from the normalized volumes as ordered stacks of \(N=5\) B-scans. For a given frame spacing \(SF\), adjacent frames in each stack were separated by \(SF\) skipped B-scans. Separate datasets were prepared for \(SF=0,5,11,\) and \(23\), representing progressively sparser sampling along the volumetric scan direction. Each volume was divided into dense nonoverlapping groups of valid sparse-frame stacks, and remaining frames were discarded if they were insufficient to form a complete five-frame stack. This procedure yielded 2550 image--label stack pairs for \(SF=0\) and \(SF=5\), and 2400 pairs for \(SF=11\) and \(SF=23\). Each sample contained the normalized image stack, corresponding binary label stack, source volume identifier, and nerve-positive or nerve-absent status.

\subsection{Training and Evaluation}
\label{sec:training_evaluation}

All segmentation models were evaluated using volume-disjoint five-fold cross-validation. The same fold assignment was used for every architecture and frame spacing, and all sparse-frame samples from a given OCT volume were assigned exclusively to either training or validation within each fold, preventing leakage from adjacent B-scans or repeated sparse stacks from the same physical volume. The compared architectures were U-Net~\cite{ronneberger2015unet}, U-Net++~\cite{zhou2020unetpp}, SegResNet~\cite{myronenko2018segresnet}, DeepLabV3+~\cite{chen2018deeplabv3plus}, Attention U-Net~\cite{oktay2018attentionunet}, TransUNet-lite adapted from TransUNet~\cite{chen2021transunet}, and NerveDetNet. Baseline widths were selected for practical and comparable segmentation capacity, with 25.95--28.75M trainable parameters across baseline networks and 13.79M for NerveDetNet.

All models were trained from scratch using the same optimizer, sampling strategy, augmentation pipeline, and primary segmentation objective. AdamW~\cite{loshchilov2019decoupled} was used with an initial learning rate of \(5\times10^{-4}\), weight decay of \(10^{-5}\), and cosine annealing to \(10^{-6}\). Each mini-batch contained 10 samples with an 8:2 ratio of nerve-positive to nerve-absent samples. Nerve-absent volumes were used as explicit negative controls and analyzed separately for false-positive behavior, while validation Dice was computed on held-out nerve-positive volumes to quantify segmentation accuracy. Training used random full-width crops with a lateral size of 512 pixels, random spatial augmentation, frame jitter, and post-normalization intensity perturbations including brightness, contrast, gamma, and additive-noise jitter. Automatic mixed precision was used for all models.

The training objective used a compound Dice--BCE segmentation loss, with an auxiliary regularization term when supported by the architecture:
\begin{equation}
\begin{aligned}
\mathcal{L} &= \mathcal{L}_{\mathrm{seg}} + \lambda_{\mathrm{aux}}\mathcal{L}_{\mathrm{aux}}, \\
\mathcal{L}_{\mathrm{seg}} &=
(1-\lambda_{\mathrm{bce}})\mathcal{L}_{\mathrm{dice}}
+ \lambda_{\mathrm{bce}}\mathcal{L}_{\mathrm{bce}}^{\epsilon}.
\end{aligned}
\label{eq:training_loss}
\end{equation}
Here, \(\mathcal{L}_{\mathrm{dice}}\) is the soft Dice loss, \(\mathcal{L}_{\mathrm{bce}}^{\epsilon}\) is binary cross-entropy with label smoothing \(\epsilon=0.05\), and \(\lambda_{\mathrm{bce}}=0.5\). The auxiliary term grouped slice-consistency, deep-supervision, size-aware foreground, and hard-sample regularization terms. It was introduced with a warm-up schedule so that early optimization was dominated by \(\mathcal{L}_{\mathrm{seg}}\); terms not supported by a given architecture were set to zero.

The final benchmark used 60 training epochs. Stochastic weight averaging (SWA)~\cite{izmailov2018averaging} was enabled after epoch 40 and updated once per epoch. Model selection was based on validation Dice, and cross-validation performance was reported as the mean and standard deviation across the five held-out folds.

\begin{figure}[hbtp]
    \centering
    \includegraphics[width=\columnwidth]{ 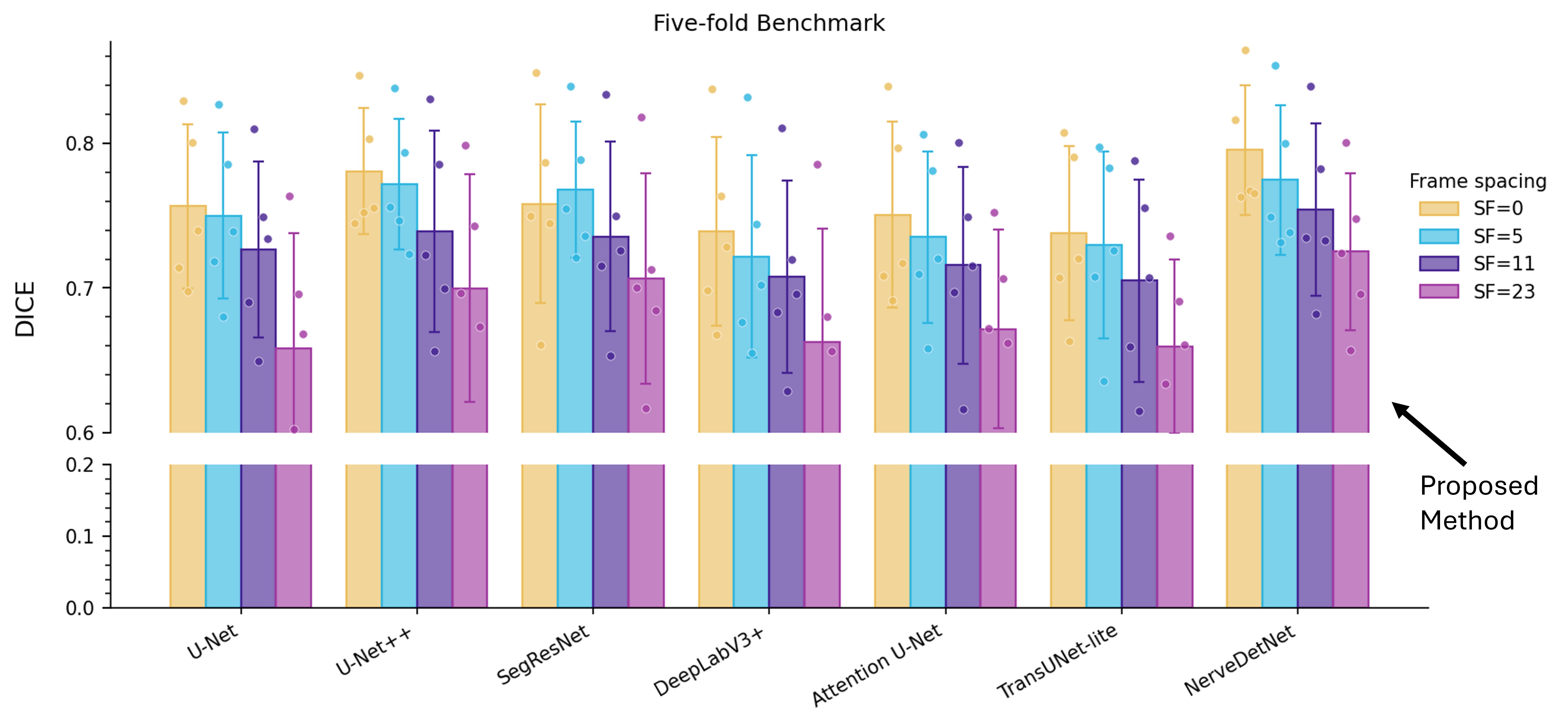}
    \caption{
    Five-fold Dice benchmark for sparse-frame OCT nerve segmentation. Bars show mean \(\pm\) standard deviation, and points show individual folds. \(SF\) denotes the skipped B-scans between frames in the five-frame input stack.
    }
    \label{fig:barchartskipframe}
\end{figure}

The primary benchmark metric was Dice on held-out sparse frame validation samples before anchor interpolation or dense-volume reconstruction. Therefore, the validation Dice in Fig.~\ref{fig:barchartskipframe} and Table~\ref{tab:detnet_dice_time} measures direct network performance under the same sparse frame input distribution used during training. Full-volume visualizations and volumetric metrics were computed separately after the anchor-interpolation procedure described in Sec.~\ref{sec:sparse_volume_reconstruction}. These metrics included intersection over union (IoU), projected clDice~\cite{shit2021cldice}, and 95th-percentile Hausdorff distance (HD95), with HD95 reported in micrometers using the OCT voxel spacing.

Runtime was measured using two complementary protocols. Network-forward time measured only model execution on preassembled GPU tensors and excluded probability-volume assembly, anchor interpolation, model loading, and file input/output. It was reported over 50 measured runs after 20 warm-up runs. Anchor-interpolated sparse-volume timing was measured after model loading and included anchor-stack construction, host--device transfer, network prediction, center-channel probability extraction, anchor-probability transfer, and linear interpolation along the B-scan direction. Model loading and file input/output were excluded. This timing was reported over 10 measured runs after 3 warm-up runs.

\begin{figure}[hbtp]
    \centering
    \includegraphics[width=\columnwidth]{ 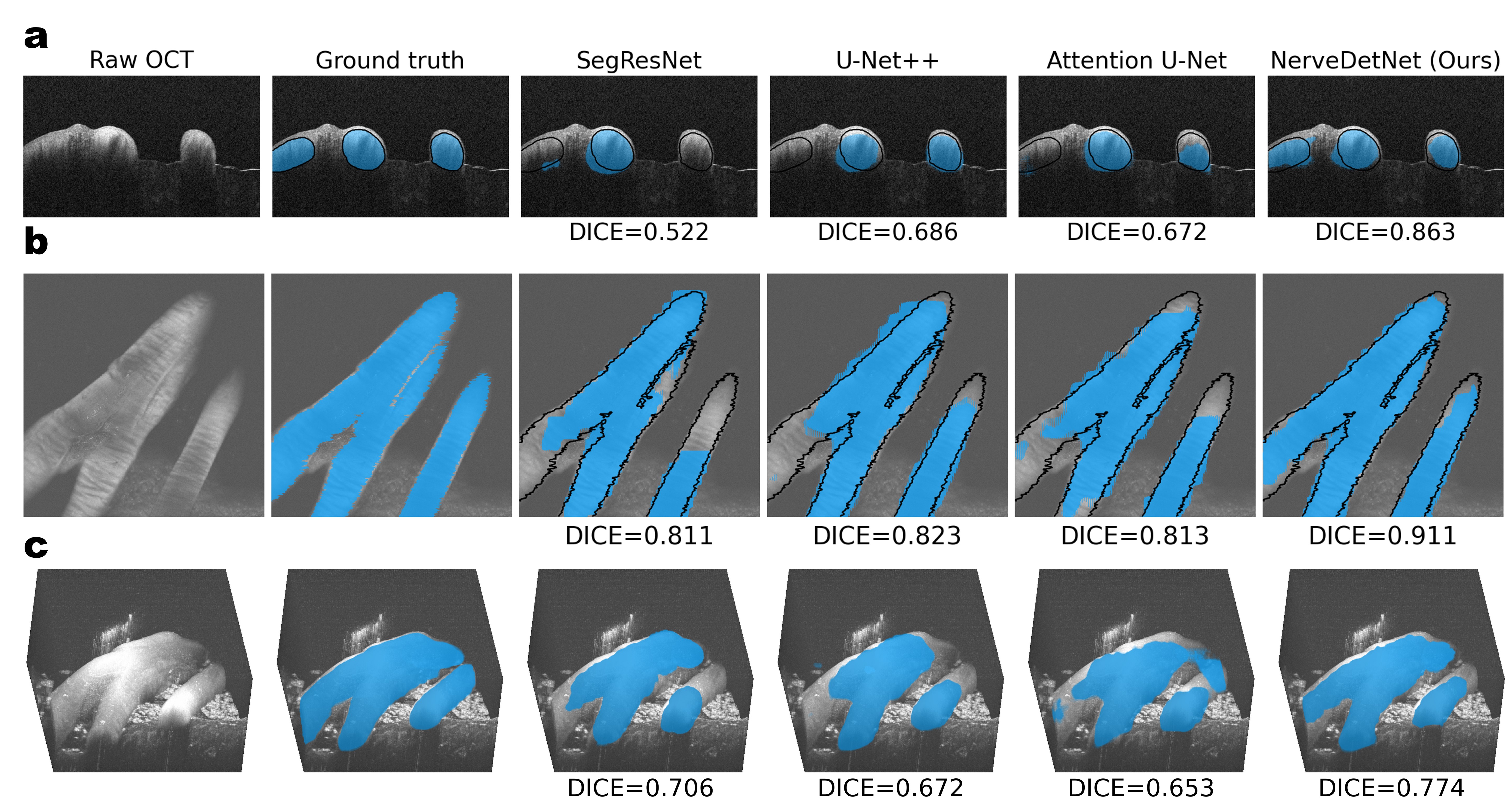}
    \caption{
    Representative validation case at \(\mathrm{SF}=23\).
    (a) B-scan predictions with slice Dice scores.
    (b) En-face projection after sparse-frame reconstruction.
    (c) 3D OCT rendering with predicted nerve masks.
    NerveDetNet shows more continuous nerve recovery.
    }
    \label{fig:nerve_case1}
\end{figure}

\begin{figure}[hbtp]
    \centering
    \includegraphics[width=\columnwidth]{ 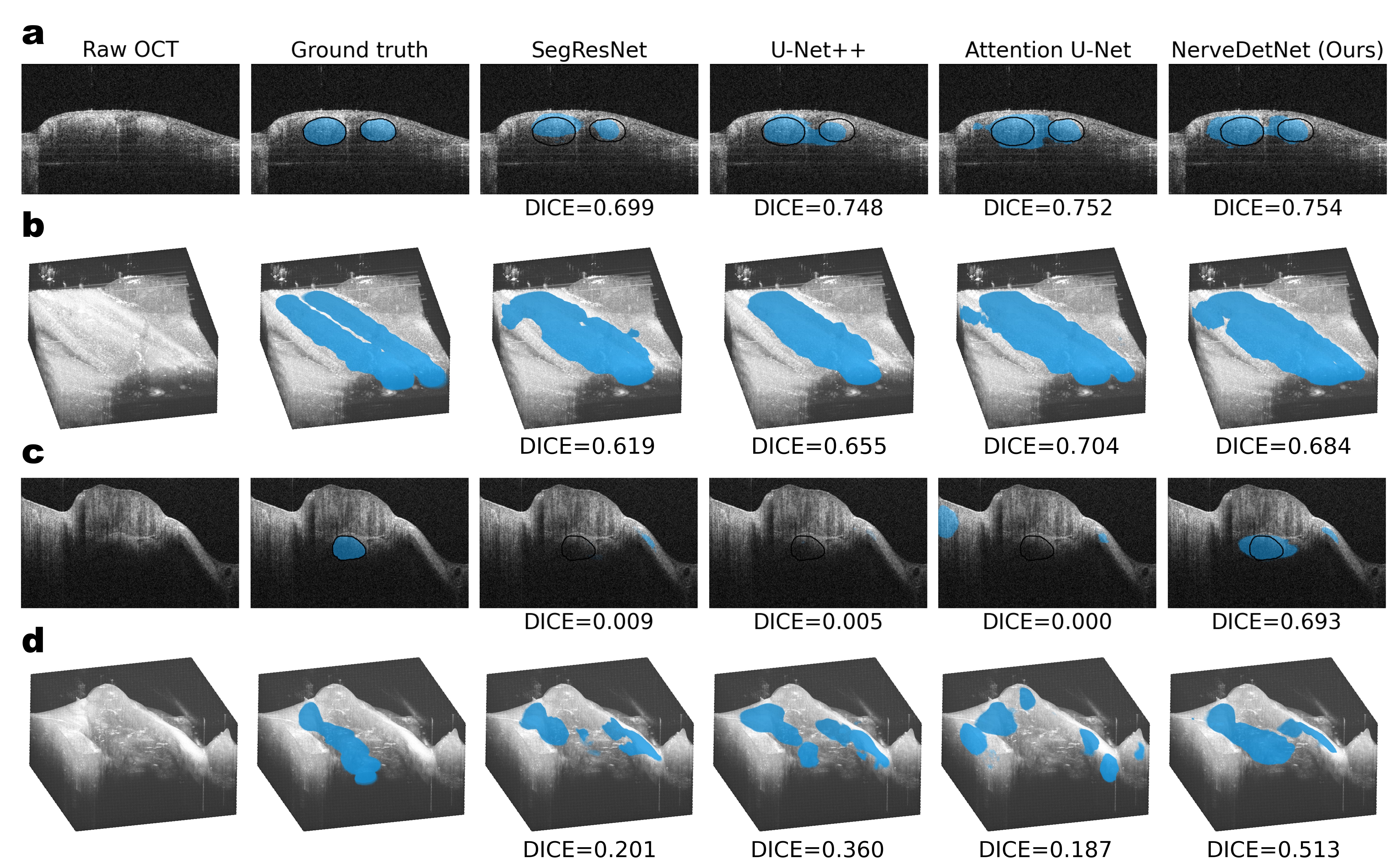}
    \caption{
    Additional shallow and deep nerve examples.
    (a,b) B-scan and 3D rendering for a shallow nerve.
    (c,d) Deep-tissue case with baseline under-segmentation.
    Dice scores are shown below predictions.
    }
    \label{fig:more_nerve_cases}
\end{figure}

\section{Results}
\label{sec:results}
\subsection{NerveDetNet Detection Performance}
We benchmarked NerveDetNet against six representative 2D segmentation architectures: U-Net, U-Net++, SegResNet, DeepLabV3+, Attention U-Net, and TransUNet-lite. All models used identical volume-disjoint five-fold cross-validation splits, OCT partitions, frame spacings, and intensity-normalization procedures. Fig.~\ref{fig:barchartskipframe} summarizes validation Dice at \(SF=0,5,11,\) and \(23\). Because Dice was computed on held-out sparse-frame samples before dense-volume interpolation, it measures direct network segmentation performance rather than post-processing effects.

NerveDetNet achieved the highest mean validation Dice at all frame spacings (Table~\ref{tab:detnet_dice_time}). The advantage was modest under dense sampling but increased with frame spacing. At \(SF=23\), NerveDetNet reached a mean Dice of \(0.725\), approximately \(0.02\) higher than the strongest baseline, SegResNet, indicating that NFCM-guided fusion is most beneficial when sparse sampling increases inter-frame displacement and nerve localization difficulty.

Runtime was measured with model loading and disk input/output excluded. For network-forward inference, NerveDetNet required \(0.229\)~s per volume, slower than DeepLabV3+, SegResNet, and U-Net, but faster than U-Net++, Attention U-Net, and TransUNet-lite (Table~\ref{tab:detnet_dice_time}). For the complete anchor-interpolated pipeline, runtime decreased with increasing \(SF\) because fewer anchor frames were evaluated (Table~\ref{tab:detnet_anchor_interp_time_by_sf}). At \(SF=23\), NerveDetNet required \(0.392\pm0.013\)~s per volume, compared with \(0.471\pm0.007\)~s for U-Net++ and \(0.400\pm0.005\)~s for Attention U-Net. Although DeepLabV3+ remained fastest, NerveDetNet provided the highest segmentation accuracy, supporting a favorable accuracy--runtime trade-off among higher-capacity baselines.

Representative reconstructed volumes at \(SF=23\) are shown in Fig.~\ref{fig:nerve_case1}. NerveDetNet more closely reproduced the annotated nerve boundary in the selected B-scan than SegResNet, U-Net++, and Attention U-Net, and preserved a more continuous trajectory in the en-face and volumetric views. Relative to the best baseline shown, NerveDetNet improved the selected B-scan Dice by \(0.177\) and volumetric Dice by \(0.068\). Additional examples in Fig.~\ref{fig:more_nerve_cases} show that, although all networks detected the dominant superficial nerve structure, NerveDetNet better preserved boundary completeness and volumetric continuity. In the deeper case, baseline models partially detected or missed the small deep nerve component, whereas NerveDetNet retained a visible response at the annotated location and preserved more of the deep nerve trajectory.

After sparse-frame reconstruction, NerveDetNet achieved the highest mean IoU and projected clDice at all frame spacings (Table~\ref{tab:detnet_extra_metrics_wide}). At \(SF=23\), it reached an IoU of \(0.664\pm0.054\), compared with \(0.643\pm0.075\) for U-Net++ and \(0.631\pm0.074\) for SegResNet, and a projected clDice of \(0.791\pm0.051\), consistent with improved preservation of elongated nerve trajectories. For HD95, NerveDetNet achieved the lowest mean value at \(SF=5\), \(SF=11\), and \(SF=23\), while SegResNet was lowest at \(SF=0\). At \(SF=23\), NerveDetNet obtained \(466.2\pm205.8~\mu\mathrm{m}\), compared with \(503.5\pm289.2~\mu\mathrm{m}\) for U-Net++ and \(512.2\pm312.2~\mu\mathrm{m}\) for SegResNet. Although U-Net++ had a slightly lower median HD95 at this spacing, NerveDetNet provided the best mean HD95 together with the highest mean IoU and projected clDice. Overall, these results show improved sparse-frame OCT nerve segmentation, particularly at larger frame spacings where missing intermediate B-scans and inter-frame displacement make nerve recovery more challenging.

\subsection{Label-Free Nerve Detection Using a Multimodal Handheld OCT Probe}

To project segmented subsurface nerves onto two-dimensional white-light surgical-view images, end-to-end label-free nerve detection combines NerveDetNet output with lateral-distortion correction, tissue-surface extraction, and cross-modal registration. We first validate these supporting steps and then demonstrate depth-resolved multimodal fusion, including nerves beneath intact overlying tissue.

\begin{figure}[htbp]
    \centering
    \includegraphics[width=7.90cm]{ 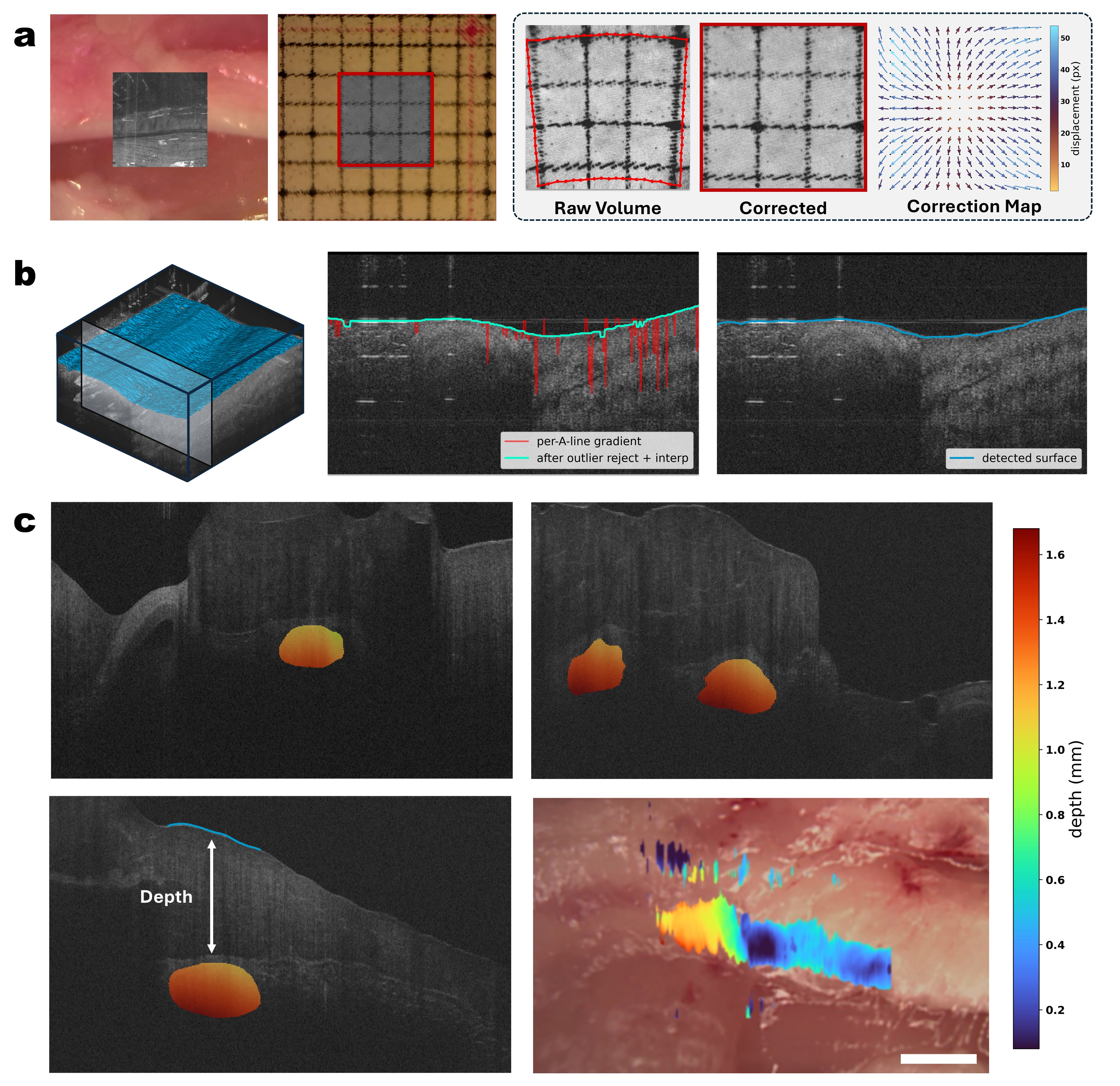}
\caption{End-to-end label-free nerve detection with the handheld multimodal OCT probe.
\textbf{(a)} OCT geometric correction and white light registration using tissue and grid targets.
\textbf{(b)} B-scan tissue-surface detection, showing gradient response, interpolated profile, and final surface.
\textbf{(c)} Depth-resolved subsurface nerve detection in representative B-scans and en-face depth fusion. Color indicates nerve depth below the tissue surface; scale bar, \(2\)~mm.}
    \label{fig:result}
\end{figure}

Geometric correction and registration are shown in Fig.~\ref{fig:result}(a). On a printed grid target, curved grid lines in the raw OCT volume were restored to straight, regularly spaced lines after thin-plate-spline correction. The displacement field increased from near zero at the field center to about \(50\)~px at the edges, consistent with MEMS scan nonlinearity, with a control-point 
\begin{table}[hbtp]
\centering
\caption{Five-fold Dice and network inference time.}
\label{tab:detnet_dice_time}
\scriptsize
\setlength{\tabcolsep}{1.8pt}
\resizebox{\columnwidth}{!}{%
\begin{tabular}{lcccc|c}
\hline
Model & SF=0 & SF=5 & SF=11 & SF=23 & Time$\downarrow$ (s/vol.) \\
\hline
U-Net & 0.756 $\pm$ 0.057 & 0.750 $\pm$ 0.057 & 0.727 $\pm$ 0.061 & 0.658 $\pm$ 0.080 & 0.098 \\
U-Net++ & 0.781 $\pm$ 0.044 & 0.772 $\pm$ 0.045 & 0.739 $\pm$ 0.069 & 0.700 $\pm$ 0.079 & 0.316 \\
SegResNet & 0.758 $\pm$ 0.069 & 0.768 $\pm$ 0.047 & 0.736 $\pm$ 0.066 & 0.707 $\pm$ 0.073 & 0.086 \\
DeepLabV3+ & 0.739 $\pm$ 0.065 & 0.722 $\pm$ 0.070 & 0.708 $\pm$ 0.066 & 0.663 $\pm$ 0.078 & \textbf{0.042} \\
Attention U-Net & 0.751 $\pm$ 0.064 & 0.735 $\pm$ 0.059 & 0.716 $\pm$ 0.068 & 0.672 $\pm$ 0.069 & 0.246 \\
TransUNet-lite & 0.738 $\pm$ 0.060 & 0.730 $\pm$ 0.065 & 0.705 $\pm$ 0.070 & 0.660 $\pm$ 0.060 & 0.238 \\
NerveDetNet & \textbf{0.795 $\pm$ 0.045} & \textbf{0.775 $\pm$ 0.052} & \textbf{0.754 $\pm$ 0.060} & \textbf{0.725 $\pm$ 0.054} & 0.229 \\
\hline
\end{tabular}%
}
\end{table}

\begin{table}[hbtp]
\centering
\caption{Anchor-interpolated inference time by frame spacing.}
\label{tab:detnet_anchor_interp_time_by_sf}
\scriptsize
\setlength{\tabcolsep}{3.0pt}
\resizebox{\columnwidth}{!}{%
\begin{tabular}{lcccc}
\hline
Model & SF=0$\downarrow$ & SF=5$\downarrow$ & SF=11$\downarrow$ & SF=23$\downarrow$ \\
\hline
U-Net & 2.151 $\pm$ 0.018 & 0.697 $\pm$ 0.012 & 0.390 $\pm$ 0.007 & 0.256 $\pm$ 0.011 \\
U-Net++ & 4.600 $\pm$ 0.018 & 1.538 $\pm$ 0.012 & 0.817 $\pm$ 0.008 & 0.471 $\pm$ 0.007 \\
SegResNet & 2.048 $\pm$ 0.032 & 0.643 $\pm$ 0.006 & 0.379 $\pm$ 0.009 & 0.242 $\pm$ 0.009 \\
DeepLabV3+ & \textbf{1.508 $\pm$ 0.025} & \textbf{0.476 $\pm$ 0.015} & \textbf{0.288 $\pm$ 0.009} & \textbf{0.214 $\pm$ 0.014} \\
Attention U-Net & 3.836 $\pm$ 0.032 & 1.273 $\pm$ 0.012 & 0.687 $\pm$ 0.010 & 0.400 $\pm$ 0.005 \\
TransUNet-lite & 3.733 $\pm$ 0.023 & 1.239 $\pm$ 0.011 & 0.673 $\pm$ 0.011 & 0.397 $\pm$ 0.008 \\
NerveDetNet & 3.672 $\pm$ 0.043 & 1.211 $\pm$ 0.013 & 0.663 $\pm$ 0.018 & 0.392 $\pm$ 0.013 \\
\hline
\end{tabular}%
}
\end{table}
\begin{table}[htbp]
\centering
\setlength{\tabcolsep}{0pt}
\caption{Volumetric segmentation results after sparse-frame reconstruction. Values are mean \(\pm\) SD, with median in parentheses.}
\label{tab:detnet_extra_metrics_wide}
\resizebox{\columnwidth}{!}{%
\begin{tabular}{@{}lcccc@{}}
\hline
Model & SF=0 & SF=5 & SF=11 & SF=23 \\
\hline
\multicolumn{5}{c}{IoU} \\
\hline
U-Net & \(.670{\pm}.066(.697)\) & \(.672{\pm}.063(.699)\) & \(.659{\pm}.063(.700)\) & \(.593{\pm}.091(.619)\) \\
U-Net++ & \(.697{\pm}.055(.707)\) & \(.674{\pm}.081(.712)\) & \(.673{\pm}.076(.697)\) & \(.643{\pm}.075(\mathbf{.693})\) \\
SegResNet & \(.674{\pm}.067(.704)\) & \(.688{\pm}.062(.702)\) & \(.668{\pm}.067(.698)\) & \(.631{\pm}.074(.673)\) \\
DeepLabV3+ & \(.644{\pm}.082(.679)\) & \(.634{\pm}.083(.653)\) & \(.637{\pm}.075(.689)\) & \(.594{\pm}.086(.631)\) \\
Attention U-Net & \(.664{\pm}.079(.688)\) & \(.655{\pm}.075(.681)\) & \(.648{\pm}.071(.681)\) & \(.603{\pm}.059(.640)\) \\
TransUNet-lite & \(.654{\pm}.072(.696)\) & \(.653{\pm}.074(.687)\) & \(.634{\pm}.076(.662)\) & \(.597{\pm}.061(.632)\) \\
NerveDetNet & \(\mathbf{.713{\pm}.054}(\mathbf{.731})\) & \(\mathbf{.696{\pm}.065}(\mathbf{.728})\) & \(\mathbf{.685{\pm}.072}(\mathbf{.700})\) & \(\mathbf{.664{\pm}.054}(.692)\) \\
\hline
\multicolumn{5}{c}{Projected clDice} \\
\hline
U-Net & \(.722{\pm}.064(.770)\) & \(.759{\pm}.060(.788)\) & \(.732{\pm}.071(.784)\) & \(.700{\pm}.115(.767)\) \\
U-Net++ & \(.799{\pm}.057(.831)\) & \(.784{\pm}.071(.832)\) & \(.802{\pm}.064(.829)\) & \(.777{\pm}.055(.810)\) \\
SegResNet & \(.778{\pm}.064(.805)\) & \(.794{\pm}.061(.821)\) & \(.767{\pm}.093(.806)\) & \(.760{\pm}.059(.801)\) \\
DeepLabV3+ & \(.776{\pm}.062(.815)\) & \(.784{\pm}.065(.814)\) & \(.806{\pm}.061(\mathbf{.864})\) & \(.786{\pm}.078(\mathbf{.829})\) \\
Attention U-Net & \(.726{\pm}.060(.774)\) & \(.722{\pm}.051(.761)\) & \(.713{\pm}.066(.766)\) & \(.699{\pm}.065(.751)\) \\
TransUNet-lite & \(.715{\pm}.055(.752)\) & \(.706{\pm}.073(.768)\) & \(.710{\pm}.099(.753)\) & \(.693{\pm}.075(.764)\) \\
NerveDetNet & \(\mathbf{.817{\pm}.069}(\mathbf{.843})\) & \(\mathbf{.833{\pm}.061}(\mathbf{.848})\) & \(\mathbf{.810{\pm}.072}(.849)\) & \(\mathbf{.791{\pm}.051}(.814)\) \\
\hline
\multicolumn{5}{c}{HD95 (\(\mu\mathrm{m}\))} \\
\hline
U-Net & \(699.8{\pm}588.5(273.9)\) & \(406.9{\pm}210.6(208.5)\) & \(610.6{\pm}453.8(256.9)\) & \(834.4{\pm}498.0(285.4)\) \\
U-Net++ & \(438.3{\pm}210.6(195.8)\) & \(570.0{\pm}444.9(215.0)\) & \(411.6{\pm}230.9(\mathbf{202.7})\) & \(503.5{\pm}289.2(\mathbf{210.3})\) \\
SegResNet & \(\mathbf{419.8{\pm}287.3}(\mathbf{186.9})\) & \(420.3{\pm}276.7(198.7)\) & \(634.1{\pm}491.5(259.1)\) & \(512.2{\pm}312.2(240.9)\) \\
DeepLabV3+ & \(544.0{\pm}265.6(265.4)\) & \(561.8{\pm}339.5(264.6)\) & \(510.8{\pm}285.3(233.4)\) & \(668.8{\pm}390.2(270.0)\) \\
Attention U-Net & \(628.0{\pm}395.3(216.6)\) & \(670.7{\pm}214.1(253.5)\) & \(738.0{\pm}432.7(234.9)\) & \(817.1{\pm}265.9(321.5)\) \\
TransUNet-lite & \(834.8{\pm}524.6(276.2)\) & \(665.8{\pm}209.4(246.5)\) & \(762.0{\pm}524.3(252.8)\) & \(917.2{\pm}446.8(315.4)\) \\
NerveDetNet & \(442.5{\pm}246.6(224.3)\) & \(\mathbf{335.8{\pm}205.5}(\mathbf{185.3})\) & \(\mathbf{385.5{\pm}194.1}(224.1)\) & \(\mathbf{466.2{\pm}205.8}(218.5)\) \\
\hline
\end{tabular}%
}
\end{table}

\noindent residual of \(0.94\)~px. After one-time pixel-level calibration, the corrected OCT en-face image and white-light view overlaid directly without per-acquisition alignment.

Tissue-surface detection is shown in Fig.~\ref{fig:result}(b). The per-A-line gradient method sharply localized the air--tissue interface in high-contrast columns but produced spurious spikes in weak columns; after outlier rejection, interpolation, fallback handling, and smoothing, a continuous surface was recovered despite specular and horizontal reflection artifacts. Stacking and filtering the per-B-scan profiles yielded the coherent 3D surface map used as the subsurface depth reference.

Combining the network-detected nerve region with the tissue surface and converting axial optical path difference using a group refractive index of \(n\approx1.4\) produced the nerve-depth map overlaid on the co-registered white-light image (Fig.~\ref{fig:result}(c)). Four deep-nerve cases are shown under a common depth scale, with nerves localized at \(0.8\)--\(1.4\)~mm below the surface and invisible in white light. The uneven upper boundaries of the detected regions reflect lateral variation in overlying-tissue thickness, which changes the round-trip optical path and shifts the apparent axial position; therefore, the depth maps follow path-induced subsurface undulations rather than a simple elliptical intensity pattern. In the case containing two nerves, both were detected independently with distinct depths. Together with the absence of detections in nerve-absent volumes (Sec.~\ref{sec:dataset_preparation}), these results indicate conditional detection based on OCT nerve structure, without a prior on nerve number, shape, or position.

The en-face fusion example shows a nerve deepening along its length as the overlying tissue thickens, with reliable detection to \(1.3\)--\(1.4\)~mm; beyond this range, no confident prediction was produced, consistent with the \(1.8\)~mm effective OCT imaging depth and the need for sufficient signal-to-noise ratio. Video~2 shows the full volume of this case. The few scattered false positives in the en-face view were spatially isolated and can be removed by connected-component filtering. Together, these experiments show that the handheld multimodal probe rectifies OCT geometry, extracts the tissue surface, registers OCT to white light, and fuses network-detected nerves into a depth-resolved display for label-free visualization of subsurface nerve presence.

\section{Discussion}
\label{sec:discussion}

NerveDetNet was developed as the deep-learning component of a label-free OCT workflow for subsurface peripheral nerve detection. Because dense 3D convolutional segmentation increases memory demand and inference latency~\cite{cicek2016threedunet,milletari2016vnet}, we adopted a sparse-frame 2.5D formulation that incorporates limited through-plane context without full 3D convolution~\cite{zhang2022bridging2d3d}. Its main architectural contribution is the NFCM-guided decoder, which introduces spatial, frame-order, and shift-correlation cues before decoder fusion. By computing these cues at reduced resolution and injecting them only along the final high-resolution decoder path, NerveDetNet explicitly models shift-tolerant cross-frame nerve consistency while limiting added computational cost.

The results indicate that this design is most beneficial under sparse sampling, where inter-frame displacement and depth-dependent OCT appearance changes are more pronounced. NerveDetNet achieved the highest average validation Dice across all evaluated frame spacings, with the clearest advantage at larger skip-frame settings. Qualitative examples further showed improved continuity in size-varying and deeper nerve regions, where weak signal and partial visibility challenge generic encoder--decoder models. These findings support OCT-specific cross-frame correlation guidance for sparse-frame nerve segmentation.

Several limitations remain. The model was trained and evaluated on ex vivo samples, and broader validation is needed across tissue preparation, probe handling, imaging depth, and acquisition settings. Deep or weakly scattering nerves remain difficult because their boundaries can be ambiguous and may overlap with surrounding connective tissue. Future work should expand annotated data across nerve sizes, depths, and tissue conditions, incorporate stronger domain-randomized augmentation, and evaluate uncertainty-aware calibration to reduce false positives in low-signal regions. The current sparse-volume reconstruction uses deterministic anchor interpolation~\cite{lehmann1999survey,thevenaz2000interpolation}, which is efficient and fair across models but assumes smooth probability variation between sampled B-scans; lightweight learned reconstruction may improve volumetric continuity at lower cost than full 3D OCT segmentation.

Beyond the network, the handheld multimodal OCT probe and confirm-then-capture workflow provide a system-level contribution. Volume scanning, CUDA-based en-face reconstruction, operator review, one-button multimodal capture, and pixel-level OCT-to-white-light registration enable freehand acquisition and direct overlay of detected nerve maps onto the tissue field. In the current experiments, nerves were detected up to \(1.3\)--\(1.4~\mathrm{mm}\) below the tissue surface, approaching the \(1.8~\mathrm{mm}\) effective OCT imaging depth. To our knowledge, this is the first demonstration of localizing peripheral nerves beneath intact, unopened tissue using a handheld OCT device rather than imaging surgically exposed nerves.

Future improvements should address field of view, acquisition speed, depth range, and tissue specificity. Wider-angle MEMS or resonant--galvo scanning could reduce the number of scans needed to survey a surgical site; MHz-class swept sources could reduce motion sensitivity; and extended-depth approaches, such as Bessel-beam illumination or computational refocusing, may increase usable detection depth. Autofluorescence should be further evaluated on fresh surgical tissue and alternative excitation wavelengths~\cite{dip2022nuvInVivo}, as it was used here only as a cross-reference because of variable contrast and fat confounding. Finally, the distinct OCT appearances of muscle, fat, blood vessels, and nerve suggest a path toward multi-tissue classification, which could reduce false positives and enable tissue-aware optical path-length conversion rather than relying on a single refractive-index assumption such as \(n\approx1.4\).

\section{Summary}
\label{sec:summary}
We presented a label-free, depth-resolved framework for peripheral nerve detection that combines a handheld multimodal OCT probe with NerveDetNet, a sparse-frame 2.5D segmentation network. The probe integrates swept-source OCT with co-registered white-light and autofluorescence imaging in a confirm-then-capture workflow, enabling freehand acquisition and pixel-level overlay of detected nerves onto the surgical field. NerveDetNet uses an NFCM-guided decoder to incorporate spatial, frame-order, and shift-tolerant correlation cues from sparsely sampled B-scans. Across all frame spacings, it achieved the highest mean Dice among the compared 2D networks, with larger gains under sparser sampling. In end-to-end experiments, the system localized and depth-resolved nerves beneath intact tissue at depths of \(1.3\)--\(1.4~\mathrm{mm}\), approaching the \(1.8~\mathrm{mm}\) effective OCT imaging depth. These results demonstrate a label-free strategy for deep-tissue nerve detection without contrast agents and support future development toward intraoperative guidance.
\section*{Conflict of Interest Statement}
R. Liang is the founder of Light Research, Inc. and LR InnovOptics, Inc. The remaining authors declare no competing interests.

\bibliographystyle{IEEEtran}
\bibliography{NerveOCT}

\end{document}